\documentclass{capstone}
\graphicspath{ {./images/} }

\usepackage[textsize=footnotesize]{todonotes}

\paperwidth=\dimexpr \paperwidth + 6cm\relax
\oddsidemargin=\dimexpr\oddsidemargin + 3cm\relax
\evensidemargin=\dimexpr\evensidemargin + 3cm\relax
\marginparwidth=\dimexpr \marginparwidth + 3cm\relax

\begin{document}

\title{ReaderQuizzer: Augmenting Research Papers with Just-In-Time Learning Questions to Facilitate Deeper Understanding}

\capstoneyear{2023}
\capstonedocument{Project 2}
\capstonesemester{Spring}

\author{Liam Richards}
\affiliation{%
  \institution{Computer Science, NYUAD}
}
\email{lrm444@nyu.edu}



\advisors{Azza Abouzied, Nancy Gleason}

\renewcommand{\shortauthors}{Richards}

\begin{abstract}
Academic reading is a key component of higher education, and serves as a basis for critical thinking, knowledge acquisition and effective communication. Research shows many students struggle with comprehension and analysis tasks with academic texts, despite the central importance of academic reading to success in higher education. Undergraduates and researchers need to internalize dense literature to scaffold their own work upon it. This reading task is time-consuming and difficult to do. Oftentimes, students struggle to actively and critically engage and as a result attain merely a cursory understanding of a paper's contents, or worse, incorrectly interpret the text. How, then, can we provide a means to more easily digest a text whilst also facilitating meaningful, critical engagement and understanding? This paper locates itself within the broader field of Human-Computer Interaction (HCI) to implement an augmented reading interface that leverages the power of ChatGPT to intelligently generate and co-locate comprehension and analysis questions in an academic paper, thereby making the paper more digestible with the end goal of facilitating deeper understanding, and developing critical reading skills. 
\end{abstract}


\maketitle

\section{Introduction} \label{introduction}
Academic reading is integral to success at higher education institutions, in part because reading academic papers is a daily task for researchers, graduate and undergraduate students alike. Literature surrounding academic reading emphasizes reading's importance and centrality to the college curriculum ~\cite{Howard, desa, AACU}. The text has to be actively and critically engaged with, as both are a crucial part of the learning process and for gaining the benefits of reading ~\cite{desa}. Academic reading is important not only because failure to complete required readings is associated with declining exam performance and research performance ~\cite{Kerr}, but also because strong reading capabilities foster effective participation in scholarly conversations ~\cite{manarin}, ultimately furthering academic research progress on the whole. 

Despite the central importance of academic reading to successful learning in higher education, struggles with academic reading comprehension are widespread, especially at the undergraduate level. While academic reading struggles are multi-faceted in nature, being influenced by a multitude of factors including "proficiency in the material’s printed language, motivation, self-regulation, academic background, self efficacy and students' academic life adjustment” ~\cite{anwar}, these struggles are largely due to both student and institutional shortcomings ~\cite{desa, Kerr}.

It remains critical, then, that readers' comprehension skills are stimulated and developed as much as possible to promote 'deep' understanding. We strive to develop a method to intelligently generate and co-locate comprehension questions in an academic paper; this would aid overall comprehensiveness. Secondly, we aim to further stimulate readers' critical engagement with a text. Hence, in the pursuit of fulfilling these goals, we (my capstone mentors and I) are motivated by the following question: \textit{Can a novel reading interface provide readers with an easier means to extract knowledge from and intellectually engage with a paper, especially those that prove to be prohibitively difficult to comprehend?}

Our proposed solution is an augmented reading interface intended for use primarily by undergraduates that incorporates just-in-time positioned learning questions in order to better facilitate deeper understanding of the text. We present ReaderQuizzer, a software prototype tool designed to support readers in their comprehension of academic texts.
 
This paper seeks to make four main contributions. First, we have characterized the problem regarding the undergraduate challenges of reading academic papers, grounded in a survey of reading comprehension literature. Second, we drew upon this existing research to provide design motivations for developing an interactive augmented reading interface that automatically generates learning questions for a given research article. The aim is to facilitate active engagement and promote deeper understanding of academic texts. Third, we developed said tool for helping both students and educators yield a deeper understanding of a academic papers' content and how they are read. This followed an iterative process, in which a pilot study was held to assess the prototype, resulting in guiding design principles and followed by a redesign of the tool. This was implemented as a web application with a \textit{Node.js} back-end that utilizes OpenAI's GPT-3.5 API ~\cite{ChatGPTAPI} to automatically generate learning questions. Finally, we provide evidence of the usefulness of the design through an evaluation study involving a number of undergraduate students using the tool. 

\subsection{Background}
The way we synthesize and acquire knowledge is rapidly evolving. There is an increasing trend to make learning visually engaging, interactive, and social; the emergence of 'social learning' platforms such as \href{https://www.perusall.com/}{Perusall}, academic collaboration networks such as \href{https://ataprod.remarqable.com/web/index.html}{Remarq}, and interactive articles published in journals such as \href{https://distill.pub/}{\textit{Distill}} and publications like \href{https://parametric.press/issue-02/}{\textit{The Parametric Press}} are testaments to this trend. As compared to static alternatives such as the Portable Document Format (PDF), which continues to dominate the digital document space online and in academia ~\cite{PDFpopularity}, interactive articles have been shown to be "more engaging, can help improve recall and learning, and attract broad readership and acclaim" ~\cite{Distill}. As a world of new "readers" inseparable from the vast and dynamic capabilities of the internet approaches, static PDFs may be becoming increasingly outmoded for learning. Learning in higher education is changing, and we aim to contribute to a movement in the academic sphere from the realm of static PDFs to a more interactive space.

\section{Literature Review} \label{literature review}
\subsection{Challenges with College Reading Comprehension} \label{challenges}
\subsubsection{Low proficiency in reading college-level texts.}

Undergraduate academic reading skills are negatively impacted by both comprehension and analysis difficulties. While Manarin et al. ~\cite{manarin} (p.47) mention the bulk of first-year college students in their Canada-based study display at least a benchmark level of comprehension proficiency, she establishes that faculty all deal with students who don’t “get” required readings: those who fail to make connections and delve deeper into a text. Additionally, college instructors observed that even when provided with study guides, college students typically read without an emphasis on comprehension ~\cite{lei}. Regarding difficulties with the texts themselves, studies such as ~\cite{Perin, shen, phakiti} determined that many students struggle with understanding vocabulary, identifying main and global ideas in text, composing summaries, and synthesizing information from multiple academic texts to produce academic writing. Furthermore, students are frequently in need of lecturers' assistance ~\cite{shen}.

Students’ skills in academic reading are influenced by a multitude of factors, including “their proficiency in the material’s printed language, motivation, self-regulation, academic background, self efficacy and students' academic life adjustment” ~\cite{anwar}. Perhaps most commonly, struggles arise from the fact many undergraduates don't learn \textit{how} to read college-level texts, which are considerably more difficult than typical high school texts; the \href{https://www.umass.edu/oapa/sites/default/files/pdf/tools/rubrics/reading_value_rubric.pdf}{AACU Reading Value Rubric} notes “even the strongest, most experienced readers making the transition from high school to college have not learned what they need to know to make sense of texts in the context of professional and academic scholarship” ~\cite{AACU}.

\subsubsection{Time restrictions and inadequate institutional support.}
Failure in learning \textit{how} to read academically comes from two main factors: for one, many undergraduates either do not have time or simply do not make time to hone their reading abilities, which is in turn fueled by the view academic reading is unessential to academic achievement ~\cite{Kerr}. Rather than reading to understand and contribute to an ongoing field of research as faculty do, undergraduates typically perform a cost-benefit analysis prior to reading, choosing only to do readings if they determine it will help them obtain a particular grade in a class or score well on an exam~\cite{Kerr}. If they attain these goals without reading, they understand that academic reading, while valuable in general, is unessential to academic achievement ~\cite{desa}. Secondly, researchers point out inadequate institutional-level reading support, as educators commonly “fail to engage in academic reading pedagogies” ~\cite{desa}. These faculty who did not teach academic reading did so either because they perceived trade-offs between academic reading and course content, or they held views of college-level reading comprehension as a static skill which undergraduates already possess. 

\subsubsection{Low Compliance in Required College Readings.}
Perhaps most importantly, the undergraduate academic context is characterized by low compliance with reading assignments. Studies show only 20–30\% of undergraduate students complete required readings ~\cite{Kerr}, and their academic performance suffers as a result ~\cite{anwar}. Undergraduates merely give a cursory and superficial reading of their assigned material ~\cite{andrianatos}, and subsequently do not take the proper time and effort to truly engage with and understand the readings, which research shows would simultaneously develop their reading skills. Reading skills are not necessarily improved at college, either; one study of 848 undergraduates found only minimal gains in reading abilities as students progressed toward college graduation ~\cite{gorzycki}. Learning opportunities are being lost, and consequently threatening the quality of higher education. This suggests better practices are needed to help undergraduates read and develop their comprehension abilities, by making reading assignments easier to digest, giving students who are less proficient more opportunities to practice and scaffold, and providing direct feedback to students' reading processes to foster engagement.

\subsection{How to Motivate Readers: Strategies to Support Reading Practices}
Given the undergraduate sphere is generally characterized by a lack of active and critical engagement with academic texts, fueled by a lack of time, motivation and inadequate support in teaching undergraduates how to read, how do we motivate readers to engage with academic texts and facilitate deeper understanding? Proposed methods to foster active engagement with academic texts tackle it at both an institutional and individual level. Firstly, some researchers place the responsibility of motivating students on the institution and educators within a higher education context, as “many studies highlight the value of instructors’ integration of explicit reading skills in regular courses, noting student improvements in academic reading and learning as a result” ~\cite{desa}. As Andianatos et al. ~\cite{andrianatos} note, “with expert guidance, students can become more knowledgeable to the extent that, like the trade masters, they have enough experience to further their own reading development.” On the other hand, some researchers underscore guiding students to a certain level of self-sufficiency is key to improving students’ learning and academic reading capabilities, particularly through "self-monitoring". Self-monitoring reading refers to the ability of a reader to remain aware of their own understanding and comprehension of the text, allowing students to identify areas that need clarification and engage with complex texts in a thoughtful and reflective manner. Research shows a positive relationship between self-monitoring and learning effectiveness: "when students are taught to ask each other a series of comprehension-monitoring questions during reading, they learn to self-monitor more often and hence learn more from what they read” ~\cite{Ambrose}. This is where our tool, ReaderQuizzer, fits in: integrating a self-monitored comprehension-question-based reading interface into a higher education setting for students could prove beneficial for overall active and critical engagement with texts, and may even help increase compliance in completing required readings.

\section{Related Work} \label{related works}

A handful of tools have been developed in the domain of improving and augmenting the process of academic reading. These tools include digital annotation tools ~\cite{azmuddin}, augmented reality ~\cite{Huisinga, Ebadi, Bursali}, and augmented reading interfaces. This paper focuses on the field of augmented reading interfaces, and I will draw upon their work and interactive systems tool design principles to guide the construction of and assessment of my own tool. 

\subsection{Reading Interfaces that Support Active Reading}
The first reading interface that proves useful is a 2021 CHI Conference paper titled 'Augmenting Scientific Papers with Just-In-Time, Position-Sensitive Definitions of Terms and Symbols' by Andrew Head et al ~\cite{scholarphi}. In this paper, the authors highlighted one aspect they believed made scientific papers particularly difficult to quickly digest: the use of nonce words. Nonce words are essentially technical terms and symbols that are unique to a paper and are not commonplace in the industry. A common difficulty researchers face with nonce words is the definition or explanatory equation lies in a previous section of the paper, forcing the reader to circle back which ultimately interrupts their reading flow and "stymies their comprehension". To combat this issue, the researchers developed \textit{ScholarPhi}, an interactive hypertext interface that helps readers efficiently access definitions of nonce words simply by clicking on them. \textit{ScholarPhi} also breaks down the definitions of symbols in equations. Additionally, the researchers determined it was important to provide the user interface for the document reader directly within a web browser without requiring a separate tool to be downloaded, hence developing the tool in both Python and React. This work proves especially useful as it serves as a good reference for the design considerations needed when developing a tool intended to improve comprehension for reading papers.

"CiteRead: Integrating Localized Citation Contexts into Scientific Paper Reading" by Rachatasumrit et al ~\cite{citeread}, a March 2022 International Conference on Intelligent User Interfaces paper again deals with the process of developing an interactive reading interface for scientific papers. This reading interface tackles the difficulty of tracking down follow-up and citing works on a topic. Google Scholar and many research paper platforms provide lists of all subsequent citing papers; unfortunately, they are disconnected from the original reference paper and provide little insight into how the paper was used. Researchers who wish to truly understand the connection must undergo the awkward process of searching for each citing paper, navigating to the reference, and frequently switch between papers. The authors address this with a tool named \textit{CiteRead}, built upon the previous \textit{ScholarPhi} reader using a React framework. I was inspired by the use of commentary in the margins, similar to the feature prevalent on Google Docs. The authors import selected important citing papers directly into the margins with 'minimal intrusiveness', facilitating connections to future research and ultimately reducing the amount of context switching, enabling smoother reading flow. Both the use of margins and a frame of reference for how to build upon already established interfaces like \textit{ScholarPhi} prove particularly useful in the development of ReaderQuizzer.

An augmented reading interface that encourages active participation from students came out in late April 2023, and is titled "ReadingQuizMaker: A Human-NLP Collaborative System that Supports Instructors to Design High-Quality Reading Quiz Questions", by Xinyi Lu et al ~\cite{ReadingQuizMaker}. As suggested by the title, ReadingQuizMaker incorporates Natural Language Processing (NLP) models into a document-viewing interface to support instructors in their design of high-quality reading comprehension question for students. Just as ReadingQuizMaker sets out to achieve, we too aim to support reading comprehension in higher education contexts. However, our tool ReaderQuizzer differs from ReadingQuizMaker in several key aspects. First, our tool focuses more on supporting students rather than instructors, allowing students themselves to generate learning questions in varying types and quantities as they see fit. Our student-driven approach gives students more autonomy and practice with self-monitoring their own learning, with the end goal of helping students' conceptual understanding of the text. Secondly, our tool is compatible with any articles that have a PDF version (not HTML), thus broadening the scope of compatible papers beyond merely publications since 2018 ~\cite{ReadingQuizMaker}. Lastly, our tool aligns with the American Association of Colleges and Universities (AACU)'s VALUE rubric that evaluates student reading comprehension capabilities, as later outlined in Section ~\ref{framework of reading comprehension}. Ultimately, these reading interfaces demonstrate a world of new "readers" is coming and a need for academia and pedagogy to adjust. Learning is changing.

\section{Design Motivations} \label{design motivations}
The design of this tool followed an iterative process. This section focuses on the insights gleaned from a preliminary pilot study and design/prototype efforts.

\subsection{Initial Prototype Efforts} \label{initial prototype}
An established method of testing and improving reading comprehension lies in learning questions \cite{Kerr}; questions that ask the reader about the text's content, and namely, to summarize their understanding of the text. Hence, a question generator that formulates learning questions based on paragraphs in a text would prove incredibly useful as a means of ensuring reading comprehension and stimulating critical reflection on the text.

As a proof-of-concept, a question generator was incorporated into an augmented PDF reading interface to test readers' understandings of the passage and prompt deeper critical reflection of the content. The question generation draws upon OpenAI's ChatGPT, an incredibly powerful generative language model, for paragraph level question generation at conveniently spaced locations. ChatGPT's initial output was promising; drawing upon a research paper on augmented reading interfaces for skimming support, I asked ChatGPT to "generate questions that test my understanding of the following text: \textit{introductory paragraph on skimming inserted here}". ChatGPT gave the following output:
\begin{itemize}
    \item What is the process of skimming and how is it commonly used by researchers?
    \item How has the shift to digital online publications impacted the practice of skimming?\item What are some challenges associated with skimming and how does it require strategic reading choices?
    \item How is skimming a skill that takes time to learn and effectively employ?
    \item What observations have been made about researchers and the potential for skimming sessions to devolve into reading sessions?
\end{itemize} 

My initial implementation of the augmented reading interface models itself after the \textit{ScholarPhi} interface built with Python and React. The tool exists web-browser-based user interface, which makes it more likely to garner widespread use than a tool that would have to be downloaded separately, inconveniencing users. Use of a familiar interface and design patterns from existing document viewers will also reduce overall adoption friction. Hence, our user interface is also implemented as an overlay atop the Mozilla Foundation's open source \textit{pdf.js} PDF reader application, as was the case with \textit{ScholarPhi} and \textit{CiteRead}. 

The process of developing the initial prototype took around 3 weeks, and was implemented in the form of a locally hosted web application. A great deal of the three weeks was spent understanding Mozilla's \textit{pdf.js} code repository and how to interface my own tool within it. I ultimately succeeded in adding my own button to \textit{pdf.js}'s toolbar to trigger the learning question generation. The main architecture of the tool is as follows: first, when the user clicks the 'Generate Learning Questions' button in the toolbar, the front-end sends a \textit{POST} request to my locally-hosted \textit{Node.js} back-end and triggers a scraping of the PDF text. Once the PDF text is scraped and added to an array, the back-end sends a request to the ChatGPT API to generate learning questions based on the text. Once the questions are generated, they are sent back and rendered on the front-end. Complications arose as at the time of constructing the prototype, January 2023, the official ChatGPT API hadn't yet been released. So, I instead relied upon an unofficial API created by Travis Fischer found on \href{https://github.com/transitive-bullshit/chatgpt-api}{his GitHub} until the ChatGPT API was released in late February. Once the API was released, I added my own personal OpenAI API key to the project; it will hence be necessary for future iterations of the tool to have users create and upload their own API keys rather than relying on mine. Overall, the tool prototype is successful at automatically generating learning questions based on PDF articles that are stored locally within the code repository. One such limitation with the initial prototype however is that the PDFs' file locations must be known and manually encoded before running the tool. 

\subsection{Prototype Pilot Study} \label{prototype pilot study}
\subsubsection{Procedure.} To better understand how an augmented reading interface might support readers in actively engaging with academic reading and to also help shape the design trajectory of \textit{ReaderQuizzer}, I conducted a small formative study with four undergraduates of varying disciplines. In the study, I met with these undergraduates in-person and had them use the prototype reading interface locally hosted on my laptop to read a yet unread academic text of their choosing; this academic text was either a required course reading or related to their major of study. During the study, they were given two tasks:
\begin{itemize}
    \item \textit{Task 1:} Participants had to read their text for a minimum of 10 minutes, and could read for longer if they felt like continuing. They were asked to engage with the questions located alongside the academic text at some point during their reading.
    \item \textit{Task 2:} Upon completing their period of reading, participants were asked a series of questions encompassing how they interacted with the comprehension questions, how it affected their reading, and whether or not they found the tool helpful and why. The questions were intentionally made broad and open-ended to ensure I did not constrain their reflection on their usage of the tool, and that feedback was as honest as possible. Questions included \textit{“Did you find the tool helpful, and why or why not? Is there anything you would change about the tool? How did you find the questions? Did you find their structure and placement helpful, and why or why not?”}
\end{itemize}

\subsubsection{Analysis and Resulting Design Motivations}
Each interview took around 25 minutes in total (10 minutes of reading and 15 minutes of post-reading discussion). Rough transcripts were created from audio recordings of the post-reading discussion, solely for reference purposes and were not shared with anyone. I analyzed the pilot study data following a qualitative approach. Throughout the analysis process, general themes were identified and refined, and relevant support utterances from participants were extracted from the transcripts. In our results, we refer to participants with the pseudonyms P1-P4. 

On the whole, response was positive to the tool, even in its initial state. The general consensus of the participants was that 1) the tool’s questions served as a mini-summary of key points in an article and 2) the tool effectively increased active engagement with the article and focus on the details. In the words of P4:
\begin{quote}
    \textit{I really like it. I think it forces me to engage a lot more with the text. At first I thought these questions looked a little basic, but especially when nowadays I feel I’m rather lazy when I skim, it’s good to have all those basic details set.}
\end{quote}
and according to P1:
\begin{quote}
   \textit{In a way I felt I was better able to understand the content of the text through the questions.}
\end{quote}

However, the tool did not address all the reading comprehension desires of the users, and good feedback was obtained regarding points to improve upon. This pilot study resulted in several design motivations that are not just specific to my tool, but can be generalized for designers and builders of intelligent reading-comprehension-testing interfaces, and are as follows:

\begin{itemize}
    \item \textbf{D1: Supply an optimal amount of questions.} Everyone seemed to have a slightly different preference for the optimal amount of questions. While P1 and P2 felt \textit{“3 to 4 questions per page is a nice amount”}, P3 disagreed, instead preferring \textit{“5 to 8 questions throughout the whole paper”}. To address this disparity, an option should be provided to allow the user to decide the number of questions.
    \item \textbf{D2: Supply different types of questions.} Participants had varying preferences for the type and scope of questions. P1 felt \textit{“more specific questions are much better than broader, vague questions. They make you think and pick out the answer from the text”}. P4, on the other hand felt \textit{“the most valuable questions were those that didn't have a direct answer in the text and forced you to think about it yourself”}. P3 had a multi-pass reading approach and wanted to generate increasingly more difficult, broad, and analytical questions with each round of reading. Thus, to support these different workflows, our design should supply users the option to generate different types of questions. 
    \item \textbf{D3: Direct users to an answer to generated questions.} Participants didn’t want to simply guess they had the correct answers to the learning questions; they wanted answers to serve as an additional verification of their understanding of the text. P2’s ideal workflow is to \textit{“mentally answer the [provided] questions, and then look at the answers and make sure that they’re aligning.”} Answers to questions serve not only as learning confirmation, but could also provide users with an additional perspective. Furthermore, P1 envisioned a future iteration of the tool that \textit{“could show where in the text each question was derived from”}, which would greatly facilitate their process of skimming the paper. Overall, providing answers to questions was recommended by users as an effective learning reinforcement mechanism. 
    \item \textbf{D4: Ensure question relevancy and minimize question repetition.} Both P1 and P3 felt questions were \textit{“irrelevant or unnecessary”} at times as they were \textit{“very similar to a preceding page”}. Although participants described skipping over these questions without a significant impact on their focus, efforts should be taken to minimize question repetition to maximize reader engagement.
    \item \textbf{D5: Support different question placement for varying article structures.} According to P3, who read a journal article subdivided nicely into related works, methodology, results, and so on: \textit{“I’d find it more helpful if the questions were structured by section rather than by page”}. If a method was introduced to either determine the article structure and generate questions accordingly or give the user the autonomy to decide, this could augment overall user-friendliness. 
    \item \textbf{D6: Support a note-taking workflow.} According to P4, \textit{“I would also love it if this tool could be integrated into the note-taking workflow, letting me take notes all in one place so I don’t have to open a separate application. I’d love to be able to type my answers and export them.”} P1 agreed, saying \textit{“I think it would be great to be able to type notes or responses to these questions as you go along.”}
\end{itemize}

Although \textit{D4}, \textit{D5}, and \textit{D6} are a bit beyond the initial scope of this capstone paper, these design points all augment the overall user experience and would make for excellent additions to a future iteration of this tool. Ultimately, design points \textit{D1-D6} prove helpful in envisioning the potential of augmented reading interfaces as an undergraduate study tool.

\subsection{Theoretical Framework of Reading Comprehension} \label{framework of reading comprehension}
To design a tool that improves reading comprehension, it is necessary to quantify reading comprehension. The American Association of Colleges and Universities (AACU) proves an authoritative source on comprehensively assessing and discussing student learning at higher education institutions. In particular, the \href{https://www.aacu.org/initiatives/value-initiative/value-rubrics/value-rubrics-reading}{AACU's Reading VALUE Rubric} articulates fundamental criteria for both understanding and assessing undergraduate reading comprehension. The AACU Reading Rubric scores reading performance in each category on a scale of 1 to 4, 1 being Benchmark and 4 being Capstone level. Undergraduates typically perform around a 1 to 2, whereas levels 3 to 4 are more typical for graduate and doctorate students. The rubric also categorizes learning into 6 main categories: Comprehension, Genres, Relationship to Text, Analysis, Interpretation, and Reader’s Voice. To aid in gauging which facets are most appropriate to target with our ReaderQuizzer tool, my capstone mentors and I used the AACU definitions provided to break down the categories as follows:

\textbf{Genre questions} are those that ask about the kind of "text" presented, with regards to how these types of texts are structured, what to expect from them, what can be done with them and how to use them.

\textbf{Relationship to Text questions} are those that ask about the set of expectations and intentions a reader brings to a particular text or set of texts. For instance, \textit{how does this text relate to the reader? How does this text impact the reader’s worldview?}

\textbf{Interpretation questions} are based around determining or construing the meaning of a text or part of a text in a particular way based on textual and contextual information. For instance, \textit{what is another way to read this text? How does this text continue dialogue within and beyond a discipline/community of readers?}

\textbf{Reader's Voice questions} pertain to how readers participate in academic discourse about texts, and question the reader's own approach to the reading. For instance, \textit{are you (the reader) merely commenting about this text, or elaborating so as to deepen ongoing discussions surrounding it?}

\textbf{Comprehension questions} are questions that are geared towards learning reinforcement, and ask the reader whether they truly understood the material presented to them and are able to reiterate what they have learned. Comprehension questions are typically fact-based whose answer can be found within the scope of the text. Example comprehension questions include \textit{“What is the key message of this section?”, “How does narrative interviewing work in the context of research studies?”}, and \textit{“How did the researchers measure success in the trial?”}.

\textbf{Analysis questions} are questions that force the reader to reflect and expand beyond the scope of the paper. These types of questions require less regurgitation and more sustained thought, forcing the reader to identify reasons or motives, identify relations across texts, and reach a conclusion. For example, analysis questions include \textit{“What are the limitations of this paper?”}, \textit{“What are the weaknesses in this writer’s argument?”}, and \textit{“How does the program in this paper compare to existing programs?”}.

The AACU defines reading as "the process of simultaneously extracting and constructing meaning through interaction and involvement with written language." Hence, drawing upon this rubric, we focus on incorporating two main facets of undergraduate reading for formulating questions within the ReaderQuizzer tool: \textbf{comprehension} and \textbf{analysis}. We chose these two categories as we felt they reflected the two principal components of effective reading, comprehension representing the process of extracting meaning from the text and analysis reflecting the process of constructing meaning beyond the scope of the text. In line with design point 2 \textit{(D2)}, our second iteration of the tool will supply both of these types of questions at the user's discretion. 

\subsection{Redesign and Implementation of Our Tool}
The user interface of ReaderQuizzer was redesigned and implemented through an iterative design process, guided by the insights gleaned from our pilot studies of initial prototypes and from appropriate design features found in related works. We indicate whenever a feature or design choice was informed by one of the six design motivations introduced in Section \ref{prototype pilot study}.

As with the prototype, we implemented the user interface as a web application, with features built atop the PDF viewing and rendering platform \textit{pdf.js}. ReaderQuizzer's user interface comprises of several features that augment the original text document through overlays, widgets and panels in the margins, largely leaving the original structure of the document unmodified, as done in \textit{ScholarPhi} and \textit{CiteRead}. To minimize the friction in adopting ReaderQuizzer, we also tried to stick to common design patterns and looks found in current PDF viewers. 

First and foremost, to address this disparity of varying question-related preferences, I opted to prompt the user with a couple customization parameters they can choose from on a drop-down menu before generating the questions. Previously, the prototype generated 3 to 4 comprehension questions per page. Now, an option should be provided to allow the user to decide the number of questions \textit{(D1)}, as well as the types of questions \textit{(D2)}. As per \textit{(D2)}, the types of questions users can generate focus on two categories outlined in Section \ref{framework of reading comprehension} and are either 1) more text-specific comprehension questions and 2) broader analysis questions. To select the number of questions to be generated, the user interacts with a slider, ranging from 1 question per page to a maximum of 10. An upper limit was decided so as to not put too much strain on my use of the OpenAI API. If a future iteration of the tool allows for users to create and rely on their own OpenAI API key instead, there will be no need for an upper bound. 

To address the desire for answers to questions, the question box from the prototype design was expanded to include a 'show answer' drop-down option; clicking on this will automatically expand the question box to display an answer to the learning question \textit{(D3)}. Answers will be obtained by expanding the prompt sent to the ChatGPT API to ask it to generate both questions and answers to the questions accordingly. For example, the prompt for generating comprehension questions is as follows:
\begin{quote}
    \textit{Write [numberOfQuestions] comprehension questions followed by answers to the questions on a new line about the following research article: [insert page text here]. Number these questions with a C (like C1, C2, etc) and output each question to a new line. Output an answer preceded with 'Answer:' to a new line after each question.}
\end{quote}

Lastly, I resolved one limitation of the initial prototype, that PDFs' file locations must be known and manually encoded before running the tool. In the redesigned version, users are able to dynamically open files of their choosing by clicking the "open file" button in the toolbar, which triggers a \textit{POST} request and uploads the PDF of their choosing to the \textit{Node.js} back-end, without needing to restart the server to render the PDF.

\begin{figure*}
    \centering
    \includegraphics[width=1\linewidth]{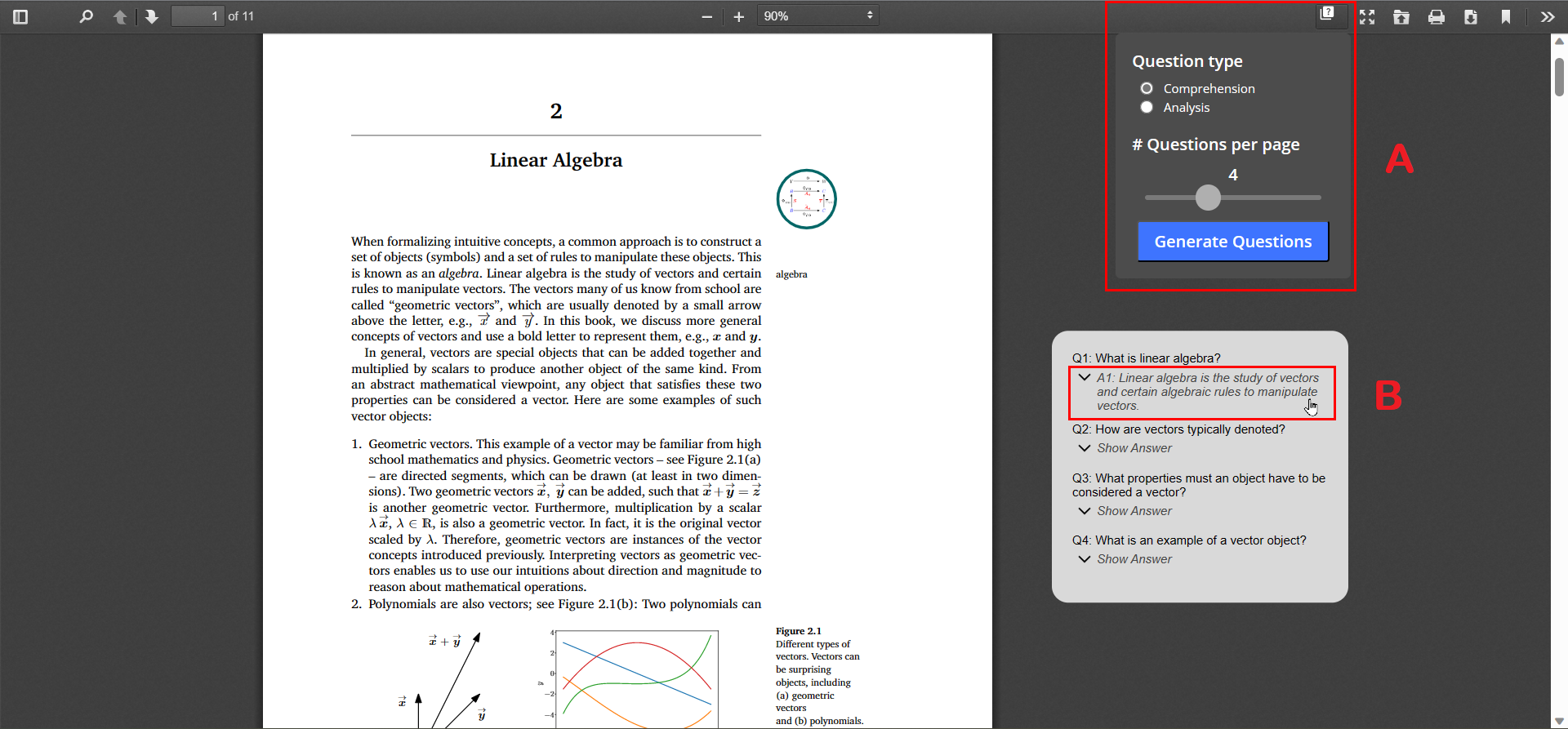}
    \caption{Redeveloped user interface of ReaderQuizzer. Learning questions are generated based on the text of each page in a paper to further stimulate reflection, knowledge acquisition and deeper engagement with the text. Guiding questions are found in the margin and are distributed throughout the text. Questions are generated by clicking the 'generate questions' icon in the toolbar. A drop-down is presented, allowing the user to select the type of question and number of questions per page, and after which the questions can be generated (A). Once generated, each page contains a question box with questions pertaining to that page's text. Answers are found below each question, and can be toggled (B).}
    \label{fig:ReaderQuizzer1}
\end{figure*}

\section{Evaluation: Usability Study}
We performed an evaluation study to understand the overall usability and usefulness of ReaderQuizzer, and collected feedback from participants on the current design, potential improvements for the system, and overall perceived usefulness as a study tool for undergraduates. We drew user feedback from the study to address the following research questions:
\begin{itemize}
    \item RQ1: Is ReaderQuizzer usable? Can students use ReaderQuizzer to generate questions that satisfy their general workflow? How do students use ReaderQuizzer for comprehending papers?
    \item RQ2: How would students perceive the questions generated by ReaderQuizzer? Are they of satisfying quality and relevance? Are they distracting? Do students find the questions generated to be helpful in promoting active and critical engagement?
    \item RQ3: What challenges do users have with the user interface and what are the general design implications for AI-based augmented reading interfaces for education?
\end{itemize}

\subsection{Participants}
We recruited participants through offline correspondences, both undergraduate classmates of mine and students of my capstone advisors. 16 undergraduate students (9 male, 7 female) participated in the study of varying class years. All participants have taken college-level courses that require readings. They are from disciplines such as computer science, mechanical engineering, economics, political science, physics, and social research and public policy. The study sessions lasted for approximately 30-40 minutes. During the study sessions, participants each selected a course-related reading text to upload to and use with \textit{ReaderQuizzer}. The text types included academic publications, textbook chapters, and their own research papers. 

\subsection{Procedure}
Participants were asked to have an academic reading of their choice ready for use prior to the evaluation session. During the session, I first got the participants’ consent to obtain and publish their feedback anonymously, and then gave a demo on how to use the ReaderQuizzer system. 

\subsubsection{Task 1: Reading Comprehension Tasks.} I prompted participants with the following scenario: Imagine you are doing this reading for a class, and you know the professor will ask you to participate in a graded discussion. Please engage with the tool’s questions as you read the paper, keeping in mind you will later be assessed on your knowledge of the paper. Afterwards, participants first had to interact with the ReaderQuizzer System, and had the tool generate a set of both analysis and comprehension questions. I briefed participants on the definitions of ‘comprehension’ and ‘analysis’ questions as outlined in Section ~\ref{framework of reading comprehension}, and encouraged them to use the tool to re-generate questions throughout their reading experience in varying amounts or question type at their discretion should it help their comprehension of the paper. At the end of the first task, participants were asked to share their thoughts on their experience. More specifically, I asked them to comment on whether the tool promoted active and critical engagement, and the challenges they experienced in the process. All participants were allowed to revisit the paper and tool at any time. All reading was done using ReaderQuizzer.

\subsubsection{Task 2: Assess the Quality of Reading Questions.} To better evaluate the questions themselves, I had participants comment on the quality of each question individually. Each question was rated on a 5 point Likert scale, prompting participants to agree or disagree with the following statement, for comprehension questions: \textit{“This question was helpful in aiding my understanding of the text, particularly in understanding main terms, identifying main and global ideas of the text, and summarizing supporting details.”} Participants were given this statement for analysis questions: \textit{“This question was helpful in aiding my understanding of the text, particularly in synthesizing information and reaching my own conclusion, understanding the strengths and weaknesses of the text, and expanding my view beyond the scope of the text.”} Participants went over a text file containing the questions and selected their answer on a Google Form. They each rated a maximum of 10 of each type of question in the interest of time.

\subsubsection{Exit Interview.} After completing the reading tasks, participants completed a System Usefulness Survey, a nine-question survey assessing ReaderQuizzer’s perceived usefulness for undergraduate reading comprehension and overall usability on a Likert scale. The survey was conducted via a Google Form. Afterwards, participants were encouraged to share any further thoughts they had on the tool, particularly regarding their overall experience using ReaderQuizzer, probing into features or interactions they liked, areas for improvement, and ideas for future augmented reading interfaces.

\subsection{Analysis}
The insights gleaned from the surveys were used to quantitatively and qualitatively evaluate the use and effectiveness of learning questions in reading comprehension. I analyzed the interview data, following the same qualitative approach outlined in the the prototype pilot study in Section \ref{prototype pilot study}. General transcripts were created from audio recordings of the post-task discussions, solely for reference purposes, and the transcripts were not shared with anyone. Emerging themes from the data were identified and refined, and relevant utterances from participants were extracted from transcripts to support our findings for each research question. In our results, we refer to participants with the pseudonyms P1-P16.

It should be noted that while all participants used \textit{ReaderQuizzer} to complete academic readings as outlined in \textit{Task 1}, not all completed \textit{Task 2} and the \textit{Exit Interview}, as participants opted to simply verbalize their thoughts and experiences with ReaderQuizzer in the interest of time. As a result, I only obtained 5 participant responses for \textit{Task 2} and 7 responses for the \textit{Exit Interview}. Aggregated results are shown in Figure \ref{fig:LikertResults} and Figure \ref{fig:LikertResults2}.

\section{Results}
We present findings corresponding to each research question.

\subsection{RQ1: All participants successfully used ReaderQuizzer to generate questions and mostly felt it improved overall reading comprehension skills.}
\begin{figure*}
    \centering
    \includegraphics[width=1\linewidth]{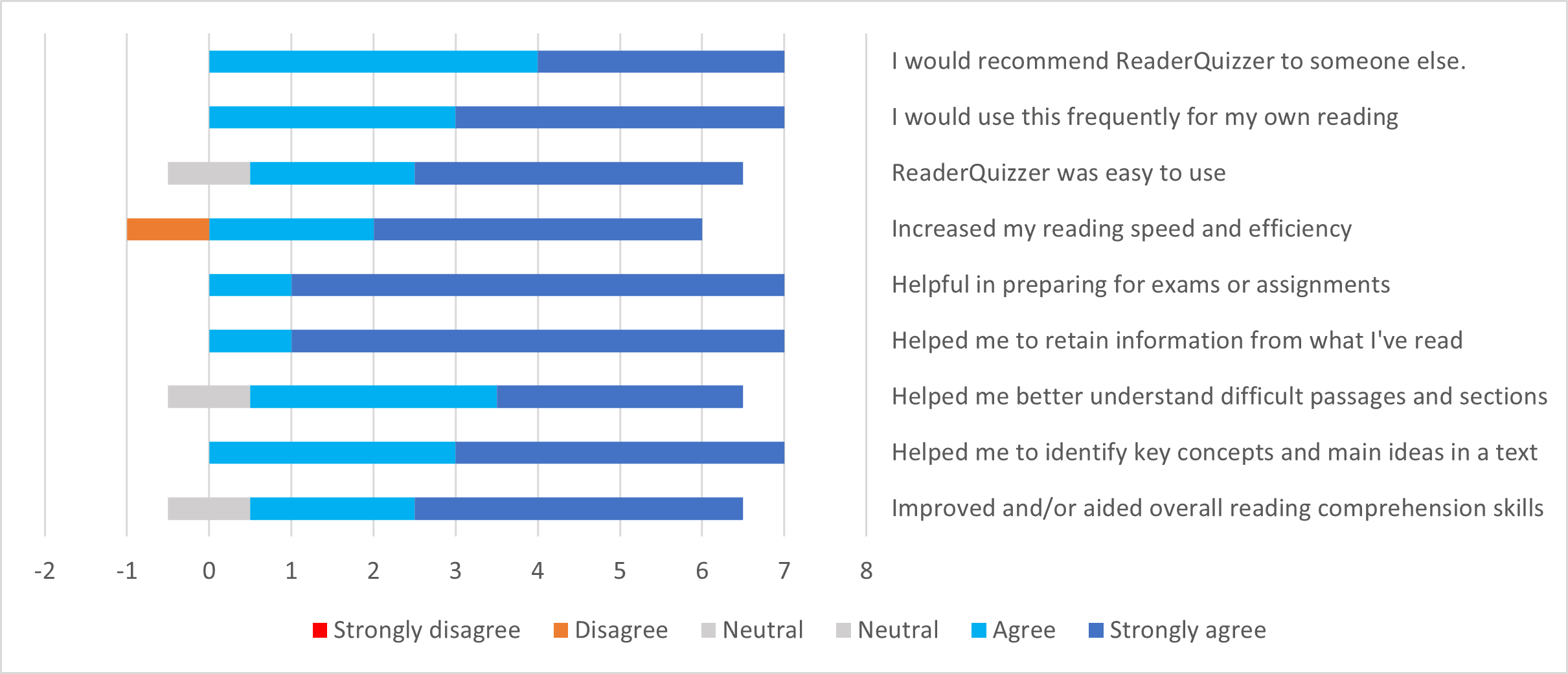}
    \caption{Distribution of responses regarding the usefulness of ReaderQuizzer in the post-experiment system usefulness survey.}
    \label{fig:LikertResults}
\end{figure*}

All users successfully uploaded their academic readings to \textit{ReaderQuizzer} and used it to generate both comprehension and analysis questions. As show in Figure \ref{fig:LikertResults}, the vast majority of participants found \textit{ReaderQuizzer} helpful in aiding their overall reading comprehension skills, identifying key concepts and main ideas, understanding difficult passages, information retention, and for exam and assignment preparation. Several participants commented on the time-saving aspect of the tool as opposed to their typical reading habits. For instance, P4 said \textit{"When I was pressed for time with a long reading assignment, ReaderQuizzer helped direct me to the most salient points, then allowed me to quiz myself on my comprehension."} According to P9, \textit{“I could see this being a very effective tool for performing literature reviews and shortening the process of research. It’s less time consuming than reading normally, as I don’t have to go through millions and millions and irrelevant details. In a typical 10-page paper, maybe only 2-3 pages of details are relevant to my research, so I like how this tool highlights relevant points and details in the question and answers.”} In these cases, participants were using the questions and answers as more of a means to summarize the text and as a result, lessen the time needed to read a text.

\begin{figure*}
    \centering
    \includegraphics[width=1\linewidth]{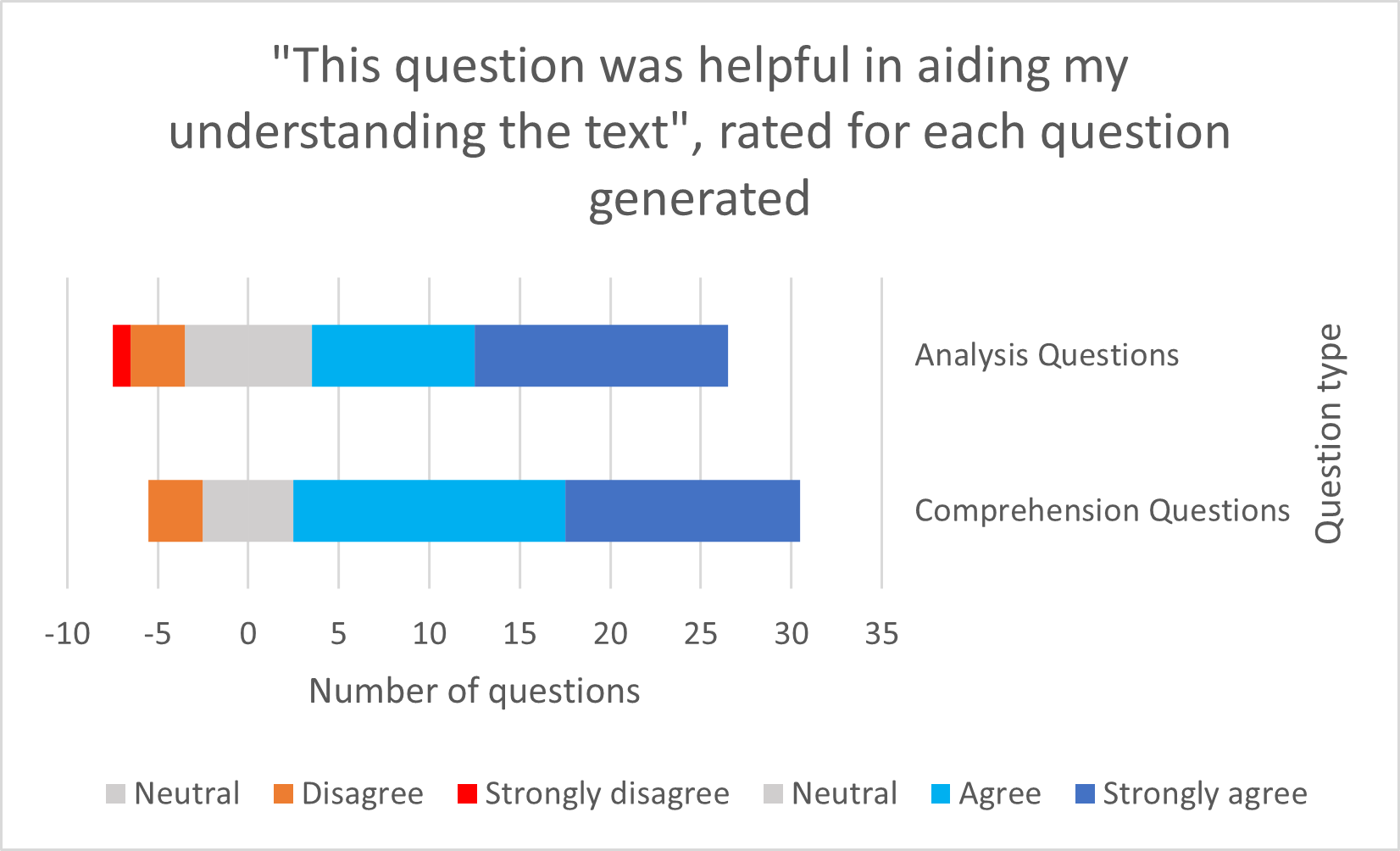}
    \caption{Distribution of responses regarding the participants' perceived quality of the questions generated by ReaderQuizzer.}
    \label{fig:LikertResults2}
\end{figure*}

\subsection{RQ2: Students find the questions and answers generated helpful in promoting active and critical engagement.}
All of the students were satisfied with the quality of learning questions that ReaderQuizzer generated, and expressed excitement that they would use ReaderQuizzer as a study tool, especially when preparing for class discussions. For instance, P8 said \textit{"I found it motivating me to do my reading, like having the question and answer format makes me feel more confident with discussing my readings to my professor or classmates. I feel like it’s more digestible this way.”}. P4 also commented on the increased motivation aspect of doing readings with ReaderQuizzer, saying \textit{"ReaderQuizzer helped me complete my reading assignment quicker and with improved comprehension. This tool made my assignment feel more approachable and helped motivate me to get it done."} P2 also said \textit{"With ReaderQuizzer I’m more actively thinking about what I’m reading, and found I’m zoning out less. I’m getting more out of the texts than I was before.”} A chart of aggregated results of participants' responses regarding the perceived quality of the learning questions is shown in Figure \ref{fig:LikertResults2}.

\subsection{RQ3: User challenge and experiences in ReaderQuizzer and design implications}
Even though participants generally found \textit{ReaderQuizzer} easy to use and helped them with motivation and active and critical engagement, they reported challenges in the process of using the tool. P3 said \textit{"The main drawback for me was that the questions did not always align with their associated page. I also found that I could not use Ctrl+F while in the tool, which was unfortunate."} P5 felt similarly, wishing that ReaderQuizzer would allow them to open multiple papers at once in several tabs, as the tool currently only supports one paper open at a time. As evidenced by these testimonies, ReaderQuizzer did not encompass all workflows. However, it was interesting to observe the different ways in which participants used the tool in ways I did not initially expect and account for. For instance, some participants plugged into ReaderQuizzer academic papers that they themselves had written with the goal of evaluating and improving upon the comprehensiveness of the paper. P10 uploaded her own paper and found ReaderQuizzer determined certain sections that were insufficient: \textit{“One of questions ReaderQuizzer generated highlighted an area of improvement for my paper. The answer for the question indicated ‘this paper is missing an answer to this question’, showing me an easy way for me to beef up my paper.”}

\section{Discussion}
In this section, we discuss limitations of our study and potential future directions. The evaluation study found that ReaderQuizzer was overall well received by our participants. Students found the system relatively easy and intuitive to use, the questions generated useful to their reading comprehension of the text, and looked forward to incorporating tools such as ReaderQuizzer regularly into their academic reading routines. Here we discuss some remaining research that could be done with \textit{ReaderQuizzer} and more broadly in the realm of augmented reading interfaces for education.

\subsection{Limitations}
One limitation of our study is its focus participant sample is small. In the future, ideally we would run a larger, longitudinal study to better understand how undergraduates use tools like ReaderQuizzer to hone their reading comprehension skills, to understand how students may develop trust with the system. Longitudinal access to ReaderQuizzer would shed insight into assessing how readings might truly use the tool "in the wild", and may also help participants discover further limitations that would preclude widespread adoption. 

Furthermore, in the evaluation study we asked users to self-report whether the tool helped them increase their reading speed and efficiency. This question was meant to serve as another indication as to whether students are likely to adopt a system like ReaderQuizzer to enhance their reading process. A more comprehensive qualitative study is needed to fully investigate the time-saving and reading efficiency aspect of ReaderQuizzer. Finally, this work focused on a student-facing evaluation study that investigated the usefulness of ReaderQuizzer for study practices and reading comprehension of required academic readings for courses. A larger-scale study is needed to understand whether use of the tool is beneficial for student performance in classroom discussions associated with the academic readings uploaded to the tool. 

\subsection{Future Work}
\subsubsection{Make ReaderQuizzer a fully-online web app to promote widespread use and adoption.} One of the biggest drawbacks of ReaderQuizzer in its current state that would prevent widespread adoption in higher education contexts is the installation process. As ReaderQuizzer is run by hosting a local Node.js server on the user's PC, Node had to be installed and environment variables declared in order for ReaderQuizzer to run properly. The proposed alternative is hosting ReaderQuizzer on a server and making it available online, likely at a site like \textit{readerquizzer.com}, where users each have a unique session with the service and can dynamically upload their articles for reading. The trade-off here is with the OpenAI API key that ReaderQuizzer relies on. As it currently stands, users may add their own API key to the ReaderQuizzer codebase so their use of the tool is directly billed to them. However, if ReaderQuizzer were hosted online, billing would be directed through 1 or several API keys of mine, so a suitable workaround with rate limiting or subscription plan must be created so that the costs of running ReaderQuizzer are met effectively.

\subsubsection{Decrease loading times.}
In ReaderQuizzer, once the user clicks "generate questions", the user needs to wait for the entire ChatGPT API call to complete before they can view the learning questions. Oftentimes this takes around 5 seconds per page, which results in minutes of waiting time for a lengthy paper. In the evaluation study, we did not observe users to be annoyed or distracted by the latency since they typically spent the time reading the original text. Perhaps there is a way the questions could be displayed as soon as they are retrieved from the API, rather than waiting for the entire API call to complete. Future work that incorporates large language models in user interfaces needs to reduce the waiting time and examine the potential effects on the overall user experience.

\subsubsection{Enhanced personalization of the reading experience.}
In addition to improving the overall speed of the interface and reduce waiting time, future reading interfaces could be better tailored to individual readers' characteristics and preferences. ReaderQuizzer currently supports 2 question types and a dynamic number of questions per page, yet future iterations could encompass multi-pass reading workflows, position questions differently based on the text type (for instance, questions solely about the introduction at large, methodology, results, and so on rather than generate questions by page). One could also tailor the experience to users of different fields, and evaluate their goals in reading the paper and the number of times the paper has been read previously.

\subsubsection{Increased engagement for students with attention deficit disorders.}
During our evaluation study, one of our participants disclosed that they have an attention deficit disorder, and that ReaderQuizzer helped increase their engagement with the text when otherwise their attention would be brought elsewhere. It lends the question, could more engaging augmented reading interfaces help students with attention deficiencies? Further studies need to be conducted regarding the efficacy of augmented reading interfaces in relation to the study habits of students with attention deficit disorders to determine whether this technology could prove beneficial. 

\section{Conclusion}
We presented \textit{ReaderQuizzer}, an augmented reading interface that automatically generates 2 types of learning questions, comprehension and analysis, based on the AACU Reading rubric ~\cite{AACU}, to support students in their reading comprehension of academic texts. In an evaluation study with 16 participants, we found that participants were satisfied with the quality of the learning questions, with a resulting increased motivation to read, comprehension of a text, and self-reported information retention, and was reported as helpful for facilitating active and critical engagement with a text. Students looked forward to incorporating tools such as \textit{ReaderQuizzer} into their studies to assist with classroom discussions and exam preparation. With the advent of interactive reading interfaces such as \textit{ReaderQuizzer}, the way reading is being conducted in academia is rapidly evolving and the learning experience is being reshaped. Based on the qualitative insights from our study, we recognize the potential of augmented reading interfaces for improved motivation in self-monitored learning, personalized learning, and increased meaningful engagement, and argue that educators and intuitions should actively adapt the way reading is being taught and assessed to embrace this upcoming era of enriched learning. 

\bibliographystyle{ACM-Reference-Format}
\bibliography{refs}

\end{document}